\documentclass[aps,prb,twocolumn,floatfix,superscriptaddress,showpacs]{revtex4-1}

\usepackage[dvipdfmx]{graphicx}
\usepackage[mathscr]{euscript}
\usepackage{amsmath,amssymb}
\usepackage{bm}
\usepackage{color}
\usepackage[vcentermath]{youngtab}
\usepackage{ulem}

\usepackage[dvipdfmx,
colorlinks=true,
urlcolor=blue,
citecolor=blue,
linkcolor=blue,
hyperfootnotes=false]{hyperref}

\def\ket#1{ | #1 \rangle }
\def\bra#1{ \langle #1 | }

\newcommand{\1}{\mbox{1}\hspace{-0.25em}\mbox{l}} %
\definecolor{darkgreen}{rgb}{0.0, 0.2, 0.13}
\definecolor{darkyellow}{rgb}{0.84, 0.77, 0.17}

\begin{document}


\title{Roles of easy-plane and easy-axis XXZ anisotropy and bond alternation on a frustrated ferromagnetic spin-$1/2$ chain}

\author{Hiroshi Ueda}
\affiliation{Computational Materials Science Research Team, RIKEN Center for Computational Science (R-CCS), Kobe, Hyogo 650-0047, Japan}
\affiliation{JST, PRESTO, Kawaguchi, Saitama, 332-0012, Japan}

\author{Shigeki Onoda}
\affiliation{Condensed Matter Theory Laboratory, RIKEN, Wako, Saitama 351-0198, Japan}
\affiliation{Quantum Matter Theory Research Team, RIKEN Center for Emergent Matter Science (CEMS), Wako, Saitama 351-0198, Japan}

\begin{abstract}
  The spin-$1/2$ Heisenberg chain with a ferromagnetic first-neighbor exchange coupling $J_1$ and an antiferromagnetic second-neighbor $J_2$ has a Haldane dimer ground state with an extremely small spin gap. Thus, the ground state is readily altered by perturbations. Here, we investigate the effects of XXZ exchange magnetic anisotropy of both the easy-axis and easy-plane types and an alternation in $J_1$ on the ground state, the spin gap, and magnetic properties of the frustrated ferromagnetic spin-$1/2$ chain. It is found that there are two distinct dimerized spin-gap phases, in one of which the spin gap and the magnetic susceptibility are extremely small around the SU(2) symmetric case and in the other they are moderately large far away from the SU(2) symmetric case. A small alternation in the amplitude of $J_1$ rapidly shortens the pitch of spin correlations towards the four-spin periodicity, as in the limit of $J_1/J_2\to0$. These effects are not sufficient to quantitatively explain overall experimentally observed magnetic properties in the quasi-one-dimensional spin-gapped magnetoelectric cuprate Rb$_2$Cu$_2$Mo$_3$O$_{12}$ that exhibits ferroelectricity stabilized by a magnetic field. Our results are also relevant to Cs$_2$Cu$_2$Mo$_3$O$_{12}$, where the ferromagnetic intrachain and antiferromagnetic interchain order has recently been found,  in a single chain level.  We also reveal the nature of symmetry-protected topological phase transitions in the model by mapping onto effective spin-1 chain models.
\end{abstract}

\maketitle
\section{Introduction}
\label{sec:intro}

The frustrated spin-$1/2$ XXZ chain with a ferromagnetic first-neighbor exchange coupling $J_1<0$ and an antiferromagnetic second-neighbor exchange coupling $J_2>0$ has attracted considerable interest both for realizing nontrivial phases, including the vector-spin-chirality ordered phase~\cite{Furukawa10} and the Haldane dimer phase~\cite{Furukawa12}, and for the relevance to quasi-one-dimensional edge-sharing multiferroic cuprates~\cite{Furukawa10}, such as A$_2$Cu$_2$Mo$_3$O$_{12}$ (A=Rb, Cs)~\cite{Hase04,Hase05,Fujimura16,Goto17,Yagi18,Hayashida19}, LiCu$_2$O$_2$~\cite{Masuda05,Park07,Seki08,Lorenz09}, SrCuO$_2$~\cite{matsuda95}, LiCuVO$_4$~\cite{Enderle05,Naito07,Yasui08,Enderle10,Mourigal11}, LiCuSbO$_4$~\cite{Dutton12}, Li$_2$CuZrO$_4$~\cite{Drechsler07}, and PbCuSO$_4$(OH)$_2$~\cite{Yasui11,Wolter12}. The Hamiltonian is given by
\begin{align}
    \mathcal{H}_{\mathrm{XXZ}} = \sum_{n=1,2}J_n \sum_{j}\sum_{\alpha=x,y,z}
    \Delta^\alpha S_j^\alpha S_{j+n}^\alpha,    
    \label{eq:H_0_XXZ}\\
    \Delta^x=\Delta^y=\Delta^{xy},
  \nonumber
\end{align}
with the spin-$1/2$ operator $\bm{S}_i$ at the site $i$.

The model has been studied intensively and extensively in the Heisenberg ($\Delta^{xy}=\Delta^{z}=1$)~\cite{Tonegawa89,Tonegawa90,Krivnov96,Itoi01,Lu06,Dmitriev07,Mahdavifar08,Kumar10,Furukawa12,Agrapidis2017,Efthimia2019} and easy-plane ($\Delta^{xy}=1$ and $0<\Delta^{z} < 1$) cases~\cite{Tonegawa90, Somma01,Jafari07,FSSO08,Sirker10,Furukawa10,FSF10,SFOF11,Furukawa12}. For $J_1/J_2\lesssim-4$, the ground state belongs to a Tomonaga-Luttinger liquid (TLL) phase with a quasi-long-range ferromagnetic order~\cite{Tonegawa90}. For $J_1/J_2\gtrsim-4$, on the other hand, the model exhibits various phases, depending on $\Delta^{z}$. The ground state belongs to the Haldane dimer phase~\cite{Furukawa12}, which has been labeled as the D$_+$ phase~\cite{UO14}, around the SU(2)-symmetric case. The subscript $+$ or $-$ denotes the relative sign of the $xy$, and $z$ components of the dimer order parameters and is opposite to the parity eigenvalue of the ground state~\cite{UO14}.
The spin gap in this D$_+$ phase is orders of magnitude smaller~\cite{Itoi01, Furukawa10, Furukawa12,Agrapidis2017,Efthimia2019} than in another D$_+$ phase realized for antiferromagnetic $J_1$~\cite{Majumdar69, Haldane82, Okamoto92, Nomura94, White96, Nersesyan98, Hikihara01,Efthimia2019}. 
With increasing easy-plane exchange anisotropy, namely, decreasing $\Delta^{z}$ from unity, a gapless vector-chiral (VC) phase appears robustly~\cite{Furukawa10}. In practice, these states are susceptible to other weak perturbations. For instance, an infinitesimally small alternation $\delta$ in the amplitude of $J_1$, as described by
\begin{align}
    \mathcal{H}_{\delta\mathrm{XXZ}} = J_1 \delta \sum_{j} (-1)^{j-1} \sum_{\alpha}\Delta^\alpha S_j^\alpha S_{j+1}^\alpha,
\label{eq:H_0_xyz_delta}
\end{align}
replaces the gapless VC phase with two topologically distinct vector-chiral dimer (VCD$_+$ and VCD$_-$) phases separated by a gapless VC phase boundary~\cite{UO13,UO14}. Unfrustrated three-dimensional interchain interactions readily leads to a long-range spiral magnetic order, as is the case in multiferroic cuprates~\cite{Furukawa10,Masuda05,Park07,Enderle05,Naito07,Yasui08,Yasui11,Wolter12}. With a further decrease in $\Delta^{z}$, there appears another dimer (D$_-$) phase, which can have a larger spin gap than in the D$_+$ phase. The coexistent phases, i.e., the VCD$_+$ and VCD$_-$ phases, survive in narrow regions even for $\delta=0$~\cite{Furukawa10,Furukawa12}. For more details, see Ref.~[\onlinecite{Furukawa12}] in the case of $\delta=0$ and Refs.~[\onlinecite{UO13,UO14}] in the case of $\delta\ne0$.

In contrast to the easy-plane case, the global phase diagram of the Hamiltonian Eq.~\eqref{eq:H_0_XXZ} in the easy-axis ($0<\Delta^{xy}<1$ and $\Delta^z=1$) case has not been fully clarified yet. It has been known that a fully polarized ferromagnetic phase (FPF) appears when $J_1/J_2<-4$~\cite{Tonegawa89, Tonegawa90}. In the Ising limit $\Delta^{xy}=0$, a first-order phase transition occurs at $J_1/J_2=-2$ from the FPF phase to an up-up-down-down (UUDD) ($\cdots \!\! \uparrow \uparrow \downarrow \downarrow \!\! \cdots$) antiferromagnetic phase~\cite{Igarashi89}. A weak inplane exchange interaction, namely, $\Delta^{xy} \ll 1$, induces a partially polarized ferromagnetic (PPF) TLL phase between the FPF and UUDD phases~\cite{Igarashi89, Tonegawa90}. 
Exact-diagonalization calculations~\cite{Tonegawa90} and a bosonization analysis~\cite{Furukawa12} have revealed that the leading correlations in this PPF TLL phase are given by multi-magnons or multi-spinons, as in the associated Heisenberg model under a magnetic field~\cite{Hikihara08,Sudan09,Heidrich-Meisner09}. However, the phase diagram with an intermediate region of $\Delta^{xy}$ remains open both with and without the bond alternation $\delta$, and its clarification is one of the two main goals of this paper. 

Although the $J_1(<0)$-$J_2(>0)$ spin-1/2 chain model should certainly be relevant to quasi-one-dimensional edge-sharing cuprates, quantitative comparisons are not necessarily easy. In particular, since the spin gap is extremely small or even vanishes near the SU(2)-symmetric case, as we have mentioned above, various perturbations may critically alter the ground state and/or excitations of the system. For instance, recent experiments combined with theoretical studies~\cite{UO18} have indicated an emergent spin-1 Haldane gap in the quasi-one-dimensional frustrated ferromagnetic spin-$1/2$ magnet Rb$_2$Cu$_2$Mo$_3$O$_{12}$~\cite{Hase04}. It has been argued that quantitative explanations of magnetic properties of the compound demand a two-leg ladder model comprising of antiferromagnetically coupled $J_1$-$J_2$ frustrated spin-1/2 chains with moderately large Dzyaloshinskii-Moriya interactions. Namely, the single frustrated ferromagnetic chain should not be sufficient. It is the other main goal of this paper to reveal the effects of various perturbations within a single $J_1$-$J_2$ chain on experimentally observable quantities. In particular, we investigate the effects of easy-plane and easy-axis exchange magnetic anisotropy and the bond alternation $\delta$ on the spin gap, the periodicity of dominant spin correlations, and the uniform magnetic susceptibility. These results will be useful for direct comparisons with experiments and indeed preclude single-chain scenarios for Rb$_2$Cu$_2$Mo$_3$O$_{12}$.

The rest of the paper is organized as follows.
In Sec.~\ref{sec:global}, we present the global phase diagram of the Hamiltonian $H={\mathcal{H}}_{\mathrm{XXZ}}+\mathcal{H}_{\delta\mathrm{XXZ}}$ and the maps of the spin gap, the periodicity of dominant spin correlations, and the uniform transverse magnetic susceptibility. Our results are consistent with the previous results in already known cases with both easy-plane~\cite{Furukawa10,Furukawa12,UO13,UO14} and easy-axis anisotropy~\cite{Igarashi89, Tonegawa89, Tonegawa90}.
In Sec.~\ref{sec:transitions}, we examine in detail phase transitions in the case of easy-axis anisotropy.
In Sec.~\ref{sec:analytic}, analytical expressions of the spin gap are derived for the FPF phase that appears with easy-axis anisotropy and for the UUDD phase in the Ising limit. We also introduce a mapping onto effective spin-1 XXZ chain models, starting from the strongly dimerized limit. On the basis of this mapping, we elucidate the nature of the symmetry protected topological (SPT) phase transition of the Gaussian universality class between (VC)D$_{\pm}$ phases and the continuous phase transition of the Ising universality class between the UUDD phase and the D$_+$ phase.
Lastly, in Sec.~\ref{sec:conclusions}, we provide discussion and the conclusions, precluding single-chain scenarios for Rb$_2$Cu$_2$Mo$_3$O$_{12}$.
Possible relevance to Cs$_2$Cu$_2$Mo$_3$O$_{12}$ is also discussed.

\section{Global ground-state phase diagram and magnetic properties}
\label{sec:global}

\begin{figure*}[t]
\begin{center}
\vspace{-6mm}
\includegraphics[width=14cm]{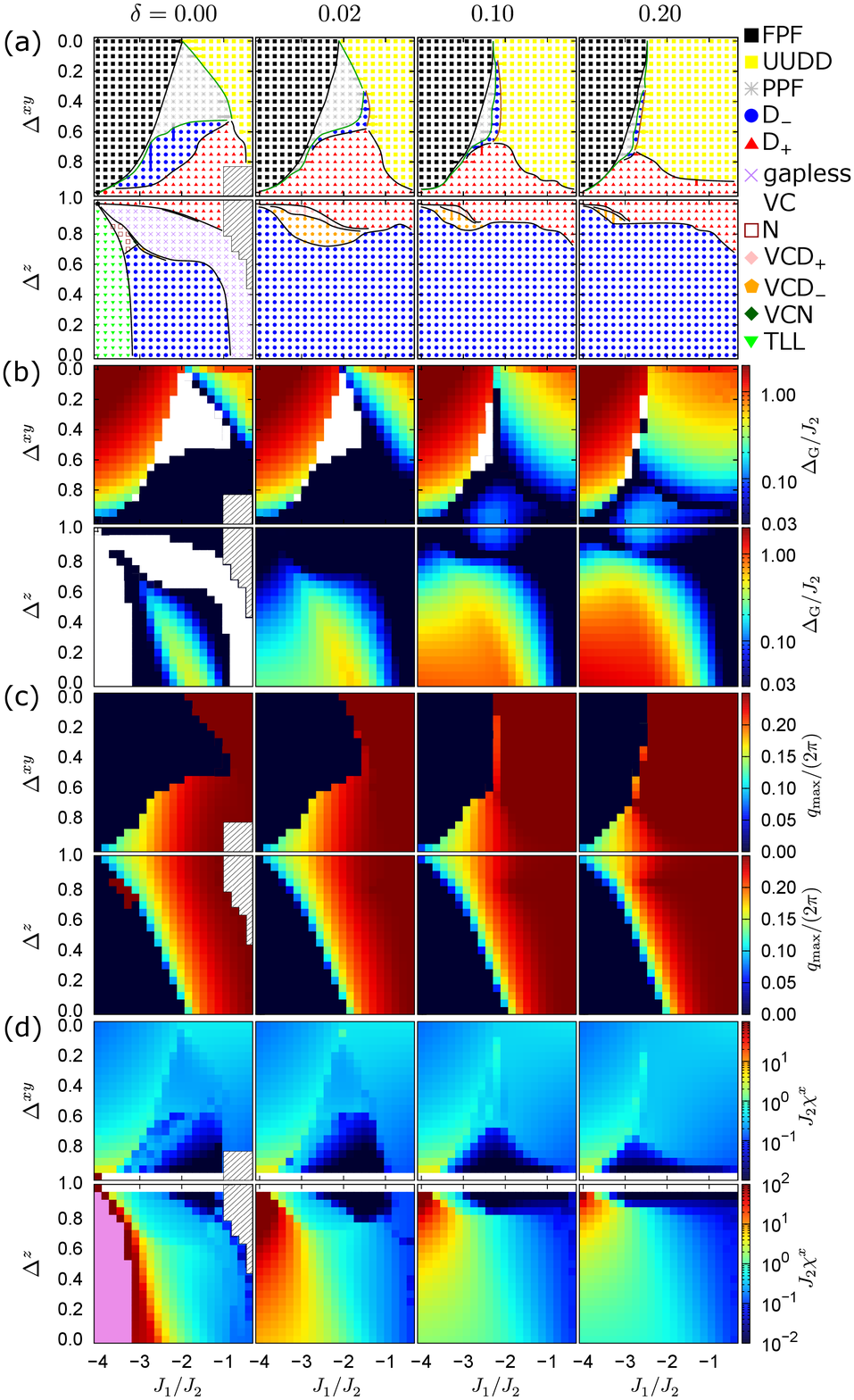}
\end{center}
\vspace{-5mm}
\caption{(a) Ground-state phase diagram, (b) spin gap $\Delta_{\mathrm{G}}$, (c) wave number $q_{\mathrm{max}}$ of the maximum spin-spin correlation function, and (d) transverse magnetic susceptibility $\chi^x$ of the bond-alternated XXZ model, $\mathcal{H}_{\rm XXZ}+\mathcal{H}_{\delta {\rm XXZ}}$. Note that in the panel (a), some symbols are overlapped with each other, forming straight lines. Thin solid lines for phase boundaries in (a) are guides to the eye. Black, green, and brown lines represent second-order, first-order, and either second-order or weakly first-order phase transitions. (See Sec.~\ref{sec:transitions}.) In the hatched areas, our iTEBD calculations using up to the matrix dimensions $300$ did not converge. In the white regions in the figure panel (b), the spin gap vanishes, i.e., $\Delta_{\mathrm{G}}=0$. In the white line respecting the SU(2) symmetry, i.e., $\Delta^{xy}=\Delta^z=1$, the magnetic susceptibility vanishes, i.e., $\chi^x=0$. In (d), $\chi^x|_{h^x \rightarrow 0}=\infty$ in the TLL phase is depicted in pink color.}
\label{fig:pd_xxz}
\end{figure*}

Figure~\ref{fig:pd_xxz} presents the main results on the global phase diagram, the spin gap $\Delta_{\mathrm{G}}$, the wave number $q_{\mathrm{max}}$ of the maximum spin-spin correlation, and the transverse magnetic susceptibility $\chi^x$ for the Hamiltonian $\mathcal{H}=\mathcal{H}_{\mathrm{XXZ}}+\mathcal{H}_{\mathrm{\delta XXZ}}$. 
The physical quantities except the spin gap are computed by the infinite-time evolving block decimation (iTEBD) method~\cite{Vidal07}. In iTEBD, we start from random complex matrix-product states (MPSs)~\cite{Affleck1988,Fannes1992,Ostlund1995,Rommer1997} with the 4-site period, in which the spatial pattern of order parameters discussed later can be embedded, and adopt the same Suzuki-Trotter decomposition~\cite{Trotter1958,Suzuki1976} as in Ref.~[\onlinecite{Furukawa10}]. 
The bond dimensions $\chi$ of the MPS are taken up to $300$, and the step size of imaginary time in Suzuki-Trotter decomposition is taken to be ${\mit\delta}\tau = 0.008/J_2$.  (See Appendices~\ref{sec:itebd} and \ref{sec:itebd_m} for numerical details of iTEBD.) The spin gap is calculated by means of the infinite-size density matrix renormalization group (iDMRG)~\cite{white92,white93,iDMRG} with $\chi$ up to 800, and we employ the procedure given in  Ref.~[\onlinecite{Efthimia2019}]. (See Appendix~\ref{sec:idmrg}.) 
Properties of each phase are summarized in Table~\ref{table:pd_xxz} and will be explained below.

\begin{table*}[t]
\caption{Summary of ground state properties for eleven phases of the Hamiltonian $H_{\rm XXZ}+H_{\delta{\rm XXZ}}$; fully polarized ferromagnetic (FPF), partially polarized ferromagnetic (PPF), N\'eel (N), up-up-down-down (UUDD), dimer D$_{\pm}$, Tomonaga-Luttinger liquid (TLL), gapless vector-chiral (VC), VC dimer (VCD$_\pm$), and VC N\'eel (VCN) phases. Five order parameters ($M$, $\mathcal{O}_N$, $\mathcal{O}_{\rm uudd}$, $D$, $\kappa$), the spin gap ($\Delta_{\rm G}$), the wave number ($q_{\rm max}$) of the maximum spin correlation functions, the dominant component $\alpha$ of the maximum spin correlation functions, and the transverse magnetic susceptibility $\chi^x$ are defined in the text. C/IC  represents commensurate/incommensurate. Values of $\Delta_{\rm G}/J_2$ and $J_2\chi^x$ refer to the cases of $\delta=0$.}
\label{table:pd_xxz}
  \begin{center}
    \begin{tabular}{lccccccccc} \hline
      Phase & $M$ & $\mathcal{O}_N$ & $\mathcal{O}_{\rm uudd}$ & $D=D^xD^z$ & $\kappa$ & $\Delta_{\rm G}/J_2$ & $q_{\rm max}$ & $\alpha$ & $J_2\chi^x$ \\ \hline \hline
     	FPF & 1/2 & 0 & 0 & 0 & 0 & $\leq-\Delta^z(J_1/J_2+1)$ & 0 & $z$ & $\gtrsim 0.2$ \\ \hline
     	PPF &  $(0,1/2)$ & 0 & 0 & \begin{tabular}{cc} 0 & ($\delta=0$) \\ Finite & ($\delta>0$) \end{tabular} & 0 & 0 & 0 & $z$ & $\gtrsim 0.2$ \\ \hline
        N & 0 & Finite & 0 & \begin{tabular}{cc} 0 & ($\delta=0$) \\ Finite & ($\delta>0$) \end{tabular} & 0 & $< 0.03$ & $\pi$ & $z$ & $\gtrsim 10$ \\ \hline
     	UUDD & 0 & 0 & Finite & Finite & 0 & $\leq \Delta^z$ & $\pi/2$ & $z$ & $\gtrsim 0.2$ \\ \hline
     	D$_+$ & 0 & 0 & 0 & $D>0$ & 0 & $< 0.03$ & C/IC & ~~~\begin{tabular}{cc} $z$ & ($\Delta^{xy}<\Delta^z=1$) \\ $x,y$ & ($\Delta^{z}<\Delta^{xy}=1$) \end{tabular} & $ \lesssim 0.1 $ \\ \hline
     	D$_-$ & 0 & 0 & 0 & $D<0$ & 0 & 
      \begin{tabular}{cc} $< 0.03$ & ($\Delta^{xy}<\Delta^z=1$) \\ $\lesssim 0.7$ & ($\Delta^{z}<\Delta^{xy}=1$) \end{tabular}     	
  & C/IC & \begin{tabular}{c} $z$ \\ $x,y$ \end{tabular} & ~~~ \begin{tabular}{c} $\gtrsim 0.1$  \\ $\gtrsim 0.3$ \end{tabular} \\ \hline
     	gapless VC & 0 & 0 & 0 & 0 & Finite & 0 & IC & $x,y$ & $\gtrsim 0.1$ \\ \hline
     	TLL & 0 & 0 & 0 & 0 & 0 & 0 & 0 & $x,y$ & $\infty$ \\ \hline
     	VCD$_+$ & 0 & 0 & 0 & $D>0$ & Finite & $< 0.03$ & IC & $x,y$ & $\gtrsim 0.1$ \\ \hline
        VCD$_-$ & 0 & 0 & 0 & $D<0$ & Finite & $< 0.03$ & IC & $x,y$ & $\gtrsim 0.7$ \\ \hline
     	VCN & 0 & Finite & 0 & 0 & Finite & $< 0.03$ & $\pi$ & $z$ & $\gtrsim 3$ \\ \hline \hline
    \end{tabular}
  \end{center}
\end{table*}

\subsection{Fully polarized ferromagnetic (FPF) phase}
\label{sec:global:FPF}

The FPF phase~\cite{Igarashi89} has the unique order parameter of the uniform magnetization
\begin{equation}
  M = \frac{1}{L}\langle S^z_{\mathrm{T}} \rangle,
  ~~S_{\mathrm{T}}^z=\sum_jS^z_j,\label{eq:M}
\end{equation}
which is pinned to 1/2, where $\langle \cdots \rangle$ denotes the ground-state expectation value with $L$ being the total number of sites. As we will show in Sec.~\ref{sec:u_phase}, the FPF phase emerges for $J_1/J_2 < 2/(1-\delta)$ in the Ising limit $\Delta^{xy}_{~}=0$ ($\Delta^z=1$). The lower bound of $|J_1/J_2|$ for the FPF phase monotonically increases with increasing $\Delta^{xy}$ for $\delta=0, 0.02, 0.1$, and $0.2$, and reaches $J_1/J_2=-2[(1+\delta)/(1-\delta)+1]$ at $\Delta^{xy}=\Delta^z=1$~\cite{Agrapidis2017}.
Early studies~\cite{Igarashi89, Tonegawa89, Tonegawa90} have shown that the instability of the FPF state occurs towards multi-magnon bound states, and that the number of magnons forming the bound state increases with increasing $\Delta^{\rm xy}$.

In the FPF phase, the spin gap $\Delta_{\rm G}$ is given by the energy difference between the lowest-energy states with $S^z_{\mathrm{T}}=\frac{L}{2}$ and $\frac{L}{2}-1$. As we will show in Sec.~\ref{sec:spin_gap},
the analytic form of $\Delta_{\mathrm{G}}$ near $\Delta^{xy}=0$ is given by $\Delta^z|J^{~}_2|(1+J^{~}_1/J^{~}_2)$ in the Ising limit and gradually decreases with increasing $\Delta^{xy}$. The value of the wave number $k_{\rm inc}$ of this first-excited state with $S^z_{\mathrm{T}}=\frac{L}{2}-1$ depends on $J_1/J_2$ as $|k_{\mathrm{inc}}| = \cos^{-1} \left[  \frac{J_1^2} {8J_2^2 } (1-\delta^2)  - \frac{  1+\delta^2 }{1-\delta^2} \right]$ for $\delta^2\leq1$ and $-1 < \frac{J_1^2} {8J_2^2 } (1-\delta^2)  - \frac{  1+\delta^2 }{ 1-\delta^2 } < 1$, as will be derived in Sec.~\ref{sec:spin_gap}.

The wave number $q_{\mathrm{max}}$
at which the spin-spin correlation
\begin{equation}
S^{\alpha}(q) = \frac{1}{L} \sum_{j,n} e^{-inq} \langle S^{\alpha}_j S^{\alpha}_{j+n} \rangle \label{eq:Sq}
\end{equation}
shows the maximum among $\alpha=x,y,z$, trivially vanishes, namely, $q_{\rm max}=0$, because of the emergence of the uniform magnetization. 

We compute the ground-state transverse magnetic susceptibility $\chi^x$ from
\begin{equation}
\chi^x = \frac{\langle S^x_{\rm T} \rangle}{L h^x}
\label{eq:chix}
\end{equation}
with the transverse magnetic field $h^x = 0.001J_2$. It decreases with decreasing $\Delta^{xy}$ in this phase. Nevertheless, it remains as large as 0.37 for $\delta=0$, 0.02, 0.1, and 0.2 with $(J_1/J_2,\Delta^{xy})=(-3.28,0.4)$, as confirmed numerically [Fig.~\ref{fig:pd_xxz}~(d)]. In the limit of $J_2=\Delta^{xy}=\delta=0$, the Hamiltonian is equivalent to one-dimensional Ferromagnetic Ising chain, then the zero-field susceptibility of the transverse Ising chain $\chi^x=\lim_{h^x\to0}\frac{\langle S^x_{\mathrm{T}}\rangle}{L h^x}$ can be obtained exactly as $J_2\chi^x=0.5$~\cite{Fisher60,Fisher63,Minami96}.

\subsection{Partially polarized ferromagnetic (PPF) phase}
\label{sec:global:PPF}

In the PPF phase~\cite{Tonegawa89}, the magnetization continuously changes within $0<M<1/2$ as a function of $J_1/J_2$ and $\Delta^{xy}$. This phase is described as a single channel TLL and thus has no spin gap. The bosonaization analysis~\cite{Furukawa12} reveals that the transverse spin correlation function $\langle S^{+}_j S^{-}_{j'}\rangle$ decays exponentially with respect to the distance, and the longitudinal spin correlation function $\langle S^{z}_j S^{z}_{j'}\rangle$ and the bond nematic correlation~\cite{Hikihara08,Sudan09} $\langle S^{+}_j S^{+}_{j+1} S^{-}_{j'}S^{-}_{j'+1}\rangle$ show power-low decays.

The dimer order parameter
\begin{equation}
D^{\alpha}= \frac{1}{L} \sum_j (-1)^{j-1} \langle S^\alpha_{j}S^\alpha_{j+1} \rangle, 
\label{eq:D}
\end{equation}
vanishes for $\delta=0$~\cite{Furukawa12}, as also confirmed by our numerical calculations. Once we introduce $\delta>0$, $M$ and $D^{\alpha}$ may appear simultaneously, because the finite $\delta$ introduces relevant cosine terms of bosonic fields into the Hamiltonian~\cite{Furukawa12} and can shift the field-locking positions. 
As shown in Fig.~\ref{fig:pd_xxz} (a), the area of PPF phase is strongly suppressed by increasing the bond alternation $\delta$ because the UUDD state has a large energy gain with respect to $\delta$, as we will show in Sec.~\ref{sec:u_phase} and Sec.~\ref{sec:effective_s1}.
The PPF phase has $q_{\rm max}=0$, as in the FPF phase.
It also has a moderately transverse magnetic susceptibility, for instance, $J_2\chi^x=0.26, 0.26, 0.42, 0.63$ for $(J_1/J_2,\Delta^{xy},\delta)=(-1.84,0.35,0), (-2.02,0.35,0.02), (-2.56,0.6,0.1)$, and $(-2.92,0.65,0.2)$, respectively [Fig.~\ref{fig:pd_xxz}~(d)].
 
\subsection{Up-up-down-down (UUDD) phase}
\label{sec:global:PPF}

The order parameter $\mathcal{O}_{\rm uudd}$ characterizing the UUDD phase is defined by
\begin{equation}
\mathcal{O}_{\rm uudd} = \frac{1}{2L} \sum_j (-1)^{j-1} \langle S^z_{2j-1}+S^z_{2j} \rangle. 
\label{eq:uudd}
\end{equation}
For $\Delta^{\rm xy}_{~}=0$, the UUDD phase emerges for $J_1/J_2 > 2/(1-\delta)$, as this criterion is an extension of the known results for $\delta=0$~\cite{Igarashi89}. (See Sec.~\ref{sec:u_phase}.) It has been known that solitons (domain walls) form propagating modes having the lowest excitation energy in the UUDD state for $\Delta^{\rm xy}_{~} \ll 1$ in the case of $\delta=0$~\cite{Igarashi89}. A large spin gap $\Delta_{\rm G}$ exists between the ground state with $S^z_{\mathrm{T}}=0$ and the lowest energy with $S^z_{\mathrm{T}}=1$ in the UUDD phase, (see Sec.~\ref{sec:u_phase}) and is enhanced by increasing $\delta$, as numerically confirmed in Fig.~\ref{fig:pd_xxz} (a) and (b). The $q_{\rm max}$ becomes $\pi/2$ to reflect the four-site periodicity of $\mathcal{O}_{\rm uudd}$. The transverse magnetic susceptibility $\chi^x$ in the UUDD is as large as $J_2\chi^x=0.38, 0.29, 0.31, 0.32$ for $(J_1/J_2,\Delta^{xy},\delta)=(-0.76,0.2,0), (-0.76,0.4,0.02), (-1.12,0.4,0.1)$, and $(-1.3,0.4,0.2)$, respectively.

\subsection{Haldane-dimer (D$^{~}_{+}$) phase}

The D$_+$ phase is characterized by $(D^x+D^y)D^z > 0$, and does not have any local magnetic order. In this phase, effective spin-1 degrees of freedom emerge on the bonds with stronger ferromagnetic correlation, forming a valence bond solid state~\cite{Affleck87} as in the spin $S=1$ Heisenberg chain~\cite{Haldane83-1,Haldane83-2}. The pattern of the dimer and thus the stronger ferromagnetic bonds has the twofold degeneracy when $\delta=0$, but is fixed by finite $\delta$ that breaks the one-site translational symmetry and doubles the unit cell. In particular, the weaker ferromagnetic bonds are entangled. This phase appears around the SU(2) symmetric case, as shown in Fig.~\ref{fig:pd_xxz}(a). It is robust against the XXZ anisotropy $\Delta^{xy}\ne\Delta^z$ and the bond alternation $\delta$, because of the $Z_2 \times Z_2$, time-reversal, and bond-center inversion symmetries protecting the topological property of the D$_{+}$ phase~\cite{UO14}.

The phase diagram and the spin gap $\Delta_{\mathrm{G}}$ in the SU(2)-symmetric case ($\Delta^{\rm xy}=\Delta^{\rm z}$) have been studied in detail~\cite{Itoi01,Furukawa12,Agrapidis2017,Efthimia2019}: $\Delta_{\mathrm{G}}$ is extremely small for $\delta=0$ and monotonically increases with increasing $|\delta|$, and the first-order phase transition occurs between the TLL and D$^{~}_{+}$ phases at $J_1/J_2=-2[(1+\delta)/(1-\delta)+1]$.

The wave number $q_{\rm max}$ of the maximum spin correlation can be incommensurate in the D$^{~}_{+}$ phase and evolves from zero towards $\pi/4$ with decreasing $|J_1/J_2|$ and increasing $\delta$ [Fig.~\ref{fig:pd_xxz} (c)]. It is notable that the transverse susceptibility is tiny only in the D$^{~}_{+}$ phase, as shown in Fig.~\ref{fig:pd_xxz} (d), and it merely amounts to 
$J_2\chi^x=7.9\times10^{-2}$ at most for the parameter line of $(J_1/J_2,\delta)=(-2.2,0)$, because the ground state is adiabatically connected to that in the SU(2) case where $\chi^x=0$, without closing the spin gap.

\subsection{Even-parity dimer (D$^{~}_{-}$) phase}

The D$^{}_{-}$ phase~\cite{Tonegawa90,Chubukov91,Somma01,Furukawa10,SFOF11,Furukawa12} is characterized by $(D^x+D^y)D^z < 0$, and also does not have any local magnetic order. This phase shares the same topological properties with the D$_+$ phase, except that the stronger ferromagnetic bonds are entangled. As in the case of the $D^{~}_{+}$ phase, the doubly-degenerated ground states emerge in the D$^{~}_{-}$ phases for $\delta=0$, and the degeneracy is lifted for $\delta\ne 0$. This phase appears on both the easy-plane and easy-axis sides.

Let us start with the easy-plane side $(\Delta^{xy}=1, \Delta^z<1$). It has already been shown that the D$_-$ phase appears in a rather large region, typically $\Delta^z\lesssim0.6$~\cite{Furukawa10}. While this phase appears in $-3\lesssim J_1/J_2 \lesssim -1$ in the case of $\delta=0$, it expands appreciably and occupies a large portion of the phase diagram in the case of $\delta\ne0$. (See Fig.~\ref{fig:pd_xxz}~(a).) Deeply inside the D$_-$ phase, the spin gap $\Delta_{\mathrm{G}}$ and the transverse magnetic susceptibility $\chi^x$ increase up to $0.4J_2$ and $0.72/J_2$ for $(J_1/J_2,\Delta^z,\delta)=(-1.66,0,0)$ and up to $1.4J_2$ and $0.88/J_2$ for $(J_1/J_2,\Delta^z,\delta)=(-4,0,0.2)$, respectively. Thus, the ground state can show a large magnetization by applying a small transverse magnetic field. 

Now we turn to the easy-axis side $(\Delta^{xy}=1, \Delta^z<1)$. Actually, the presence of the D$_-$ phase has for the first time been uncovered in the easy-axis case. As shown in Fig.~\ref{fig:pd_xxz}, it appears in a narrow region surrounded by the PPF phase, the D$_+$ phase, and the UUDD phase. With increasing $|\delta|$, the region quickly narrows.
The spin gap $\Delta_{\mathrm{G}}$ is not larger than $0.03J_2$, since the phase is rather narrow and sandwiched by a gapless PPF phase and a critical phase boundary with the D$_+$ phase [Fig.~\ref{fig:pd_xxz}~(a)]. Although the spin gap can be small as in the D$_+$ phase, the transverse magnetic susceptibility $J_2\chi^x$ is typically an order of magnitude larger ($\sim0.2$ for $(J_1/J_2,\Delta^{xy},\delta)=(-2.2,0.75,0)$ ) than in the D$_+$ phase.

Namely, on both the easy-plane and easy-axis sides, in general, the observation on the transverse magnetic susceptibility provides a key to discriminate the D$_-$ phase from the D$_+$ phase in experimental observations.

On both the easy-plane and easy-axis sides, the behavior of the wave number $q_{\rm max}$ of the maximum spin correlation is similar to that in the D$_+$ phase: it can be incommensurate and evolves from zero towards $\pi/4$ with decreasing $|J_1/J_2|$ and increasing $\delta$ [Fig.~\ref{fig:pd_xxz} (c)]. 

\subsection{Gapless vector-chiral (VC) phase}
The gapless VC phase is characterized by a long-range order of the uniform vector-chirality~\cite{Villain78}
\begin{equation}
\kappa= \frac{1}{L} \sum_{j} \langle [{\bm S}_j \times {\bm S}_{j+1}]^z \rangle. 
\label{eq:kappa}
\end{equation}
and a single-channel TLL showing a quasi-long-range inplane spiral spin correlation~\cite{Nersesyan98}. This phase occupies a wide region of the phase diagram with easy-plane anisotropy $(\Delta^{xy}=1,\Delta^z<1)$ in the case of $\delta=0$ [Fig.~\ref{fig:pd_xxz}~(a)]~\cite{Furukawa10,Furukawa12}. Then, the gapless VC phase is immediately replaced with (vector-chiral) dimer phases by introducing nonzero $\delta$~\cite{UO13,UO14}. The wave number $q_{\mathrm{max}}$ of the maximum spin correlations evolves from $\sim0.7$ to $\pi/2$ for $\Delta^{z}=0.85$ as $|J_1|/J_2$ decreases~\cite{Furukawa10,Furukawa12}, as shown in Fig.~\ref{fig:pd_xxz}~(c). Reflecting the gapless nature, this phase shows a large transverse magnetic susceptibility up to $J_2\chi^x=2.0$ for $(J_1/J_2,\Delta^z)=(-3.28,0.85)$. 

\subsection{Vector-chiral dimer (VCD$_\pm$) phases}

The VCD$_\pm$ phases have the relative sign $\pm$ of the dimer order parameters $D^x$ and $D^z$. They appear as coexistent phases of the vector-chirality order and the dimer order on the easy-plane side~\cite{Furukawa10,Furukawa12,UO13,UO14}. In the case of $\delta=0$, they are restricted to narrow regions sandwiched by the gapless VC phase and the D$_\pm$ phases. Turning on finite $\delta\ne0$ immediately replaces the gapless VC phase with the VCD$_\pm$ phases, except at the boundary of the VCD$_\pm$ phases. Then, with increasing $\delta$, the areas of the VCD$_\pm$ phases gradually decrease.
Behaviors of $\Delta_{\mathrm{G}}$, $q_{\mathrm{max}}$, and $\chi^x$ in the VCD$_\pm$ phases are similar to those in the D$_\pm$ phases, except that $J_2\chi^x$ in the VCD$_+$ phase can be slightly larger ($\gtrsim0.1$) than in the D$_+$ phase.

\subsection{Tomonaga-Luttinger liquid (TLL) phase}

The TLL phase in the case of easy-plane anisotropy with $\delta=0$ [Fig.~\ref{fig:pd_xxz}~(a)] has already been investigated intensively~\cite{Tonegawa90,Chubukov91,Somma01,Furukawa08,Sirker10,Furukawa10,Furukawa12} and does not have any long-range order, for instance, of $M$, $D^{\alpha}$, $\mathcal{O}_{\rm uudd}$, $\kappa$, and $\mathcal{O}_N$, but shows a quasi-long-range transverse ferromagnetic order and has gapless excitations [Fig.~\ref{fig:pd_xxz}~(b)]. The maximum peak position $q_{\rm max}$ is always zero, as shown in Fig.~\ref{fig:pd_xxz}~(c). Reflecting the quasi-long-range order, the transverse magnetic susceptibility $\chi^x$ diverges [Fig.~\ref{fig:pd_xxz}~(d)]. This TLL phase is immediately replaced with dimer phases by introducing finite $\delta$.

\subsection{N\'{e}el (N) phase}
The N phase~\cite{FSF10} is characterized by the  staggered magnetization
\begin{equation}
\mathcal{O}_N = \frac{1}{L} \sum_j (-1)^{j-1} \langle S^z_j \rangle. \label{eq:N1}
\end{equation}
The spin gap $\Delta_{\mathrm{G}}$ is small and not larger than $0.03J_2$, since the phase is narrow and surrounded by the gapless VC phase, the TLL phase, and the D$_-$ phase [Fig.~\ref{fig:pd_xxz}~(a)]. This phase has a large transverse magnetic susceptibility up to $2\times10^2$ for $(J_1/J_2,\Delta^z)=(-3.64,0.85)$. 
Because of the N\'eel order, the maximum spin correlation occurs at $q_{\mathrm{max}}=\pi$ in the $z$ component of spins, while the short-range spin correlations in the $xy$ component change from  commensurate ($q=0$) to incommensurate ($q>0$) inside this phase~\cite{Bursill95,Nomura05,FSF10}.
This N phase is replaced with the (chiral) dimer (VC)D$_\pm$ phases by finite $\delta$, as is clear from Fig.~\ref{fig:pd_xxz}(a).

\subsection{Vector-chiral Neel (VCN) phase}
From the bosonization analysis and numerical calculations in Ref.~[\onlinecite{Furukawa12,UO13}], the vector-chirality order and the N\'eel order can emerge simultaneously near the phase boundary among VC, N, D$_{\pm}$, and the TLL phases. This VCN phase has an extremely small spin gap $\Delta_{\mathrm{G}} < 0.03J_2$ and a commensurate pitch of $q_{\rm max}=\pi$~[Fig.~\ref{fig:pd_xxz}~(b) and (c)]. The behavior of $\chi^x$ in the VCN phase is similar to that in the N\'{e}el phase [Fig.~\ref{fig:pd_xxz}~(d)]. We did not observe the VCN phase for $\delta \geq 0.02$. 

\section{Phase transitions on the easy-axis side}
\label{sec:transitions}

\begin{figure*}[htb]
\begin{center}
\includegraphics[width=16cm]{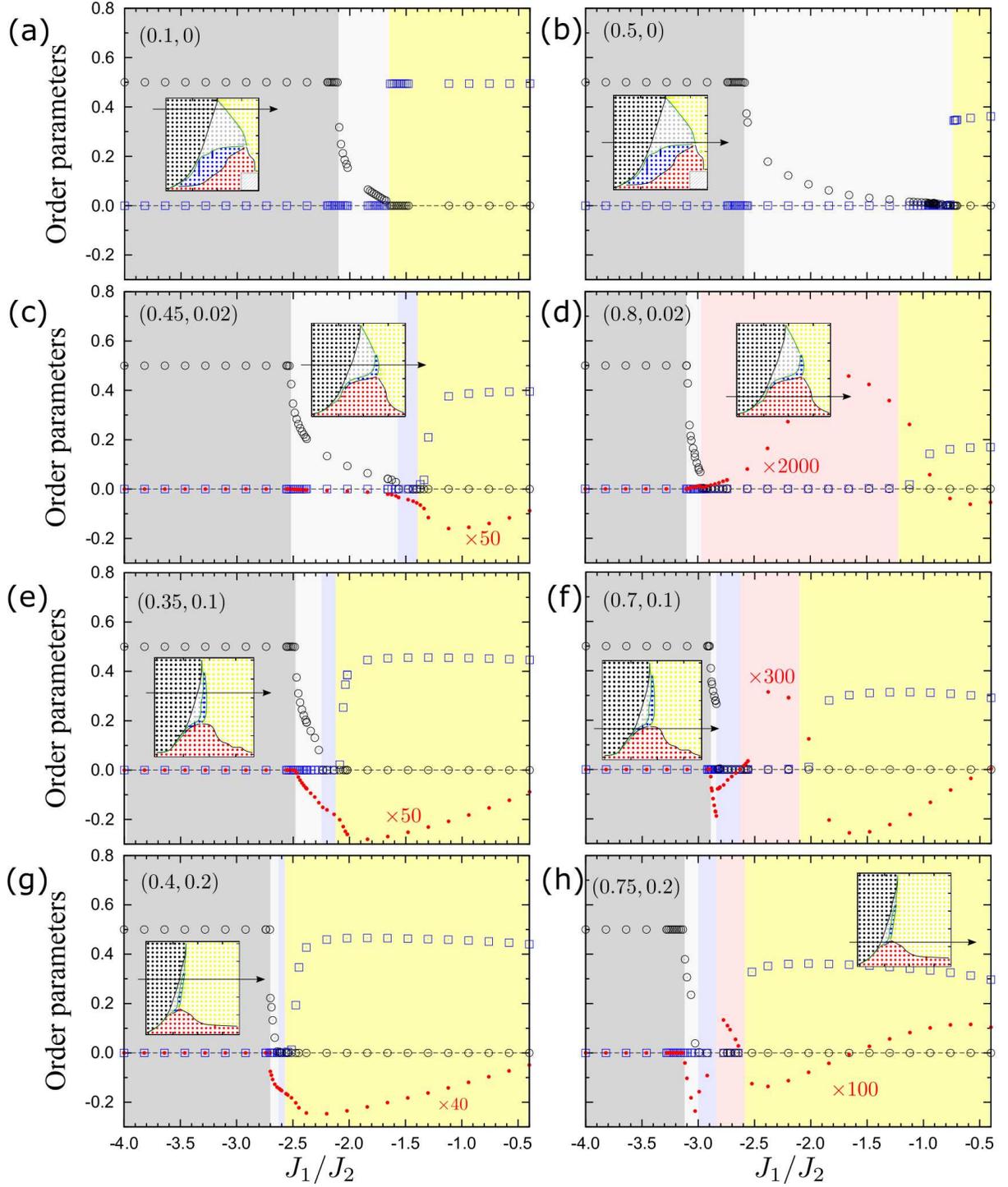}
\end{center}
\caption{Order parameters ($M: \bigcirc$, $\mathcal{O}_{\rm uudd}: \textcolor{blue}{\Box}$, $(D^x+D^y)D^z: \textcolor{red}{\bullet}$) as functions of $J_1/J_2$ for $\Delta^z = 1$. Parameter sets $(\Delta^{xy},\delta)$ for panels (a)-(h) are shown at the left top of each panels. The regions filled in dirk gray, light gray, red, blue, and yellow colors correspond to the FPF, PPF, D$_+$, D$_-$, and UUDD phases, respectively.}
\label{fig:PF_D-_U}
\end{figure*}

\begin{figure}[htb]
\begin{center}
\includegraphics[width=8cm]{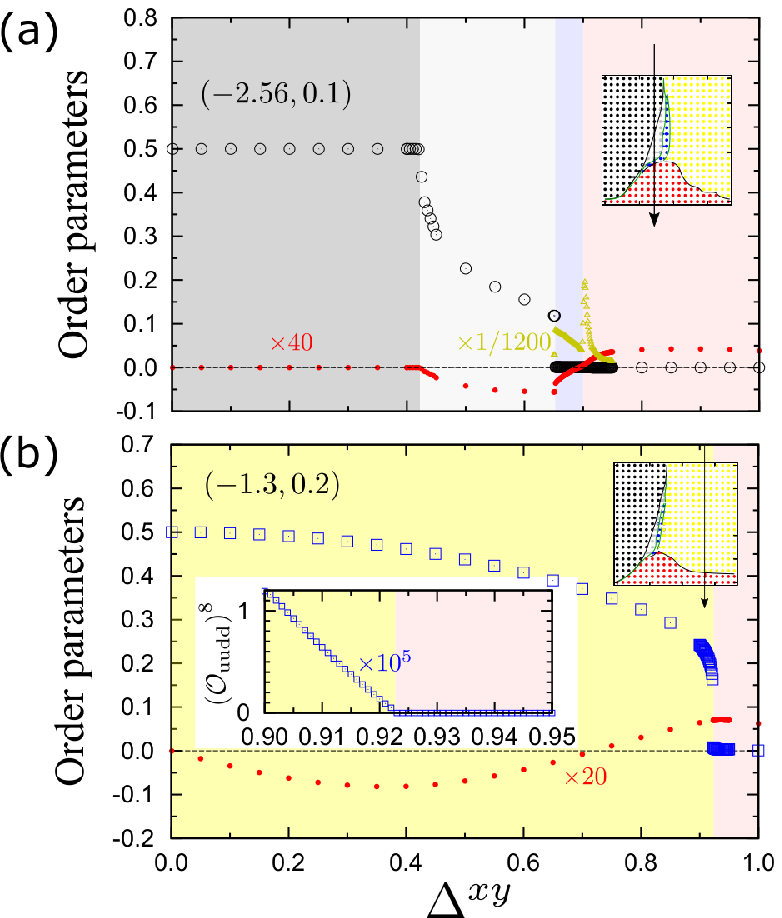}
\end{center}
\caption{Order parameters ($M: \bigcirc$, $\mathcal{O}_{\rm uudd}: \textcolor{blue}{\Box}$, $(D^x+D^y)D^z: \textcolor{red}{\bullet}$) and correlation length ($\xi:\textcolor{darkyellow}{\triangle}$) as functions of $\Delta^{xy}$ for $\Delta^z = 1$. A magnified view of $(\mathcal{O}_{\rm uudd})^8$ around $\Delta^{xy}=0.92$ is shown in the inset of panel (b). }
\label{fig:PF_D-_D+}
\end{figure}

This section is devoted to a numerically precise determination of phase boundaries in the phase diagrams on the easy-axis side [see upper panels of Fig.~\ref{fig:pd_xxz}~(a)]  and to a clarification of the nature of the phase transitions. For this purpose, we performed iTEBD~\cite{Vidal07} and/or iDMRG~\cite{iDMRG} calculations along several typical vertical and horizontal lines in the phase diagrams. In Fig.~\ref{fig:PF_D-_U}, the results of relevant order parameters are shown as functions of $J_1/J_2$ for $(\Delta^{xy},\delta)=(0.1,0)$ [Fig.~\ref{fig:PF_D-_U}~(a)], $(0.5,0)$ [Fig.~\ref{fig:PF_D-_U}~(b)], $(0.45,0.02)$ [Fig.~\ref{fig:PF_D-_U}~(c)], $(0.8,0.02)$ [Fig.~\ref{fig:PF_D-_U}~(d)], $(0.35,0.1)$ [Fig.~\ref{fig:PF_D-_U}~(e)], $(0.7,0.1)$ [Fig.~\ref{fig:PF_D-_U}~(f)], $(0.4,0.2)$ [Fig.~\ref{fig:PF_D-_U}~(g)], and $(0.75,0.2)$ [Fig.~\ref{fig:PF_D-_U}~(h)]. In Fig.~\ref{fig:PF_D-_D+}, they are shown as functions of $\Delta^{xy}$ for $(J_1/J_2,\delta)=(-2.56,0.1)$ [Fig.~\ref{fig:PF_D-_D+}~(a)] and $(-1.3,0.2)$ [Fig.~\ref{fig:PF_D-_D+}~(b)]. 
In the following, we will explain the results separately for each of the phase transitions.

\subsection{FPF--UUDD transition}
\label{sec:transitions:FPF-UUDD}
  It has already been known that a first-order phase transition occurs at $J_1/J_2=-2$ between the FPF and UUDD phases in the case of $\Delta^{xy}=\delta=0$~\cite{Igarashi89}.
 The $\delta$ dependence of the FPF--UUDD phase transition point is exactly obtained in the Ising limit as 
\begin{equation}
\frac{J_1}{J_2}= - \frac{2}{1-|\delta|}~{\rm for~} 0 \leq |\delta| \leq 1,
\label{eq:F_U_Ising}
\end{equation}
from a crossing of the energy levels of the FPF state $(\cdots \uparrow \uparrow \uparrow \uparrow \cdots)$ and the UUDD state $(\cdots \uparrow \uparrow \downarrow \downarrow \cdots)$, as we will explain in detail in Sec.~\ref{sec:u_phase}. Equation (\ref{eq:F_U_Ising}) indicates that with an increase in the bond alternation $|\delta|$, the UUDD phase expands while the FPF phase narrows. On the basis of our numerical results shown in the upper panels of Fig.~\ref{fig:pd_xxz}(a), this first-order phase transition is likely to survive a finite but small transverse interaction up to $\Delta^{xy}\sim\mathcal{O}(\delta)$, though massive calculations are required for confirming the conjecture. With further increasing $\Delta^{xy}$ within $\Delta^{xy} \ll 1$ , the direct FPF--UUDD phase transition disappears and the FPF and UUDD phases are intervened by the PPF phase.

\subsection{FPF--PPF transition}
\label{sec:transitions:FPF-PPF}
  Phase transitions between the FPF and PPF phases are observed from a change in the magnetization $M$ in all the panels of Figs.~\ref{fig:PF_D-_U} and in Fig.~\ref{fig:PF_D-_D+} (a). In the case of $\delta=0$, this transition has been described as a continuous phase transition associated with the condensation of bound multi-magnons~\cite{Igarashi89,Tonegawa89,Tonegawa90, Furukawa12}. This mechanism also holds when $\delta\ne0$. Note also that a similar mechanism holds in the spin $S=1/2$ XXZ model around the saturated magnetic field~\cite{Hikihara08,Sudan09,Heidrich-Meisner09}. 
  For $\delta=0$, our results on the FPF--PPF transition line shown in the upper leftmost panel of Fig.~\ref{fig:pd_xxz}(a) is consistent with previous exact-diagonalizaiton results~\cite{Tonegawa89,Tonegawa90}.

\subsection{PPF--UUDD transition}
\label{sec:transitions:PPF-UUDD}
  The PPF--UUDD phase transition [Figs.~\ref{fig:PF_D-_U}(a)-(b)] is a first-order phase transition. This is confirmed by the discontinuity in the relevant order parameter $\mathcal{O}^{~}_{\rm uudd}$ at the PPF--UUDD transition, as shown in Fig.~\ref{fig:PF_D-_U}(a). As the PPF phase narrows with an increase in $|\delta|$, the direct PPF--UUDD phase transition is also shortened. The D$_-$ phase eventually intervenes the PPF and UUDD phases.  [See the phase diagrams of Fig.~\ref{fig:pd_xxz}(a)].

\subsection{PPF--D$_-$ transition}
\label{sec:transitions:PPF-D_-}
  The PPF--D$_-$ phase transition [Figs.~\ref{fig:PF_D-_U}(c),(e)-(h) and Fig.~\ref{fig:PF_D-_D+}(a)] is also a first-order phase transition. This is confirmed by the discontinuity of the magnetization $M$ at the PPF--D$_-$ transition, as shown in Fig.~\ref{fig:PF_D-_D+}(a).

\subsection{D$_-$--UUDD transition}
\label{sec:transitions:D_--UUDD}
The D$_-$--UUDD phase transition [Figs.~\ref{fig:PF_D-_U}(c),(e), and (g)] emerges in the case of $|\delta| > 0$. The UUDD order parameter $\mathcal{O}_{\rm uudd}$ rapidly drops towards the D$_-$--UUDD phase transition with either a second-order phase transition or a weakly first-order. Note that the D$_-$ phase and the UUDD phase are corresponds to Large-$D$ phase and N\'{e}el phase in an effective $S=1$ Hamiltonian for $|\delta| \sim 1$, as we will show in Sec.~\ref{sec:effective_s1}. The Large-$D$--N\'{e}el phase transition in the effective model is known as a first-order phase transition~\cite{chen_prb_2003}. As long as the analogy holds, this phase transition could be of the first order.
 
\subsection{D$_+$--UUDD transition}
\label{sec:transitions:D_+-UUDD}
The mechanism of the D$_+$--UUDD phase transition [Fig.~\ref{fig:PF_D-_U}(d), (f) and (h), and Fig.~\ref{fig:PF_D-_D+}(b)] is also understood by the analysis of the effective $S=1$ Hamiltonian for $|\delta| \sim 1$ in Sec.~\ref{sec:effective_s1}. Actually, the D$_+$--UUDD transition corresponds to the Haldane--N\'{e}el transition~\cite{chen_prb_2003} in the effective $S=1$ Hamiltonian. Thus, it could belong to the Ising universality class, as long as the analogy holds. Indeed,  the order parameter $\mathcal{O}_{\rm uudd}$ shows the Ising critical behavior $(\Delta^{xy}_{\rm c} - \Delta^{xy})^{1/8}$, as shown in the inset of Fig.~\ref{fig:PF_D-_D+}(b).

\subsection{D$_-$--D$_+$ transition}
\label{sec:tansitions:D_--D_+}
The D$_-$--D$_+$ phase transition [Fig.~\ref{fig:PF_D-_U}(f) and (h), and Fig.~\ref{fig:PF_D-_D+}(a)] has the same mechanism as that on the easy-plane side, and is a second-order phase transition~\cite{Furukawa12,UO13}. This phase transition is associated with the sign change of $D^{x}=D^{y}$ ($D^z$) on the easy-axis (-plane) side. A bosonization analysis is also available~\cite{Furukawa12,UO13}. At the transition, there is a critical divergence of the correlation length of the matrix-product state (MPS), as shown in Fig.~\ref{fig:PF_D-_D+}(a). This points to a continuous phase transition. From the analysis of an effective $S=1$ Hamiltonian for $|\delta| \sim 1$, we show in Sec.~\ref{sec:effective_s1} that this D$_-$--D$_+$ phase transition belongs to the Gaussian universality class, and is a canonical symmetry protected topological (SPT) phase transition protected by the time-reversal symmetry, bond-inversion symmetry, and $Z_2 \times Z_2$ symmetry~\cite{Pollmann10,Pollmann12}.

\subsection{FPF--D$_+$ transition}
  It has been reported that a first-order phase transition occurs between the FPF phase and D$_+$ phase at $J_1/J_2=-2[(1+\delta)/(1-\delta)+1]$ in the SU(2) limit~\cite{Agrapidis2017}. At a glance of Fig.~\ref{fig:pd_xxz}(a), it is expected that this direct FPF--$D_+$ phase transition survives tiny easy-axis anisotropy, though more intensive calculations are required for the confirmation. With increasing $\Delta^{xy}$, this transition eventually bifurcates into the FPF--D$_-$ and D$_-$--D$_+$ transitions.

\section{Analytic solutions} \label{sec:analytic}
\subsection{Spin gap in the FPF phase} \label{sec:spin_gap}

Here, we assume that the number $L$ of sites in the system is even, $L=2n$, when the bond alternation $\delta$ is finite. However, the final results on the spin gap holds even when $\delta=0$ and $L$ is odd.

The FPF ground state for even $L$ is written as
\begin{equation}
\ket{{\rm F}}=\otimes_{l=1}^{n} \ket{\uparrow \uparrow}.
\end{equation}
The ground state energy is given by
\begin{eqnarray}
  E_{\rm F}=\bra{{\rm F}} H \ket{{\rm F}}
& = & \frac{n\Delta^z}{2}(J_1+J_2).
\label{eq:E_F}
\end{eqnarray}
The excited states specified by the $z$ component $\sum_i S^z = n-1 $ of the total spin and the wave number $ k \in \{0, 2 \pi / n, \cdots, 2 \pi (n-1) / n \} $ can be expressed as 
\begin{align}
  \ket{v_k} = \sum_{i=1,2} c_{ik} \ket{v_{ik}}, 
\end{align}
with coefficients $c_{ik}$ satisfying $\sum_{i} |c_{ik}|^2 = 1$, where
\begin{eqnarray}
\ket{v_{1k}} & = & \frac{1}{\sqrt{n}} \sum_{j=0}^{n-1} e^{ikj} \hat{T}^{2j} \ket{\uparrow \downarrow} 
\otimes_{l=1}^{n-1} \ket{\uparrow \uparrow}, \\
\ket{v_{2k}} & = & \frac{1}{\sqrt{n}} \sum_{j=0}^{n-1} e^{ikj} \hat{T}^{2j} \ket{\downarrow \uparrow}  
\otimes_{l=1}^{n-1} \ket{\uparrow \uparrow},
\end{eqnarray}
form  a set of orthonormal bases. We have also introduced a translation operator $\hat{T}$ through
\begin{align}
\hat{T} \ket{\uparrow \downarrow}\otimes_{l=1}^{n-1} \ket{\uparrow \uparrow} = \ket{\uparrow \uparrow} \otimes \ket{ \downarrow \uparrow}\otimes_{l=1}^{n-2} \ket{\uparrow \uparrow}, \nonumber
\end{align}

Since tha Hamiltonian $H$ commutes with $\hat{T}^2$, we obtain
\begin{eqnarray}
H \ket{v_{1k}} & = & \frac{1}{\sqrt{n}} \sum_{j=0}^{n-1} e^{ikj} \hat{T}^{2j} H \ket{\uparrow \downarrow}
\otimes_{l=1}^{n-1} \ket{\uparrow \uparrow}
, \nonumber \\ \label{eq:Hv1_1}\\
H \ket{\uparrow \downarrow}
\otimes_{l=1}^{n-1} \ket{\uparrow \uparrow}
& = & (N-2)\frac{\Delta^z(J_1+J_2)}{2}\ket{\uparrow \downarrow}
\otimes_{l=1}^{n-1} \ket{\uparrow \uparrow}
\nonumber \\
& & + \frac{J_1\Delta^{xy}(1+\delta)}{2} \ket{\downarrow \uparrow}
\otimes_{l=1}^{n-1} \ket{\uparrow \uparrow}
\nonumber \\
& & + \frac{J_1\Delta^{xy}(1-\delta)}{2} \hat{T}^2 \ket{\downarrow \uparrow}
\otimes_{l=1}^{n-1} \ket{\uparrow \uparrow}
\nonumber \\
& & + \frac{J_2\Delta^{xy}}{2} (\hat{T}^2 + \hat{T}^{-2})\ket{\uparrow \downarrow}
\otimes_{l=1}^{n-1} \ket{\uparrow \uparrow}
. \nonumber\\ \label{eq:Hv1_2}
\end{eqnarray}
We can compute $H \ket{v_2}$ in the same manner as Eq.(\ref{eq:Hv1_1}) and Eq.(\ref{eq:Hv1_2}), except that $\ket{\uparrow \downarrow}$ and $\ket{\downarrow\uparrow }$ are swapped and $\hat{T}^2$ in the third line of Eq.~\eqref{eq:Hv1_2} is replaced with $\hat{T}^{-2}$.

Using Eqs.~\eqref{eq:Hv1_1} and \eqref{eq:Hv1_2} as well as the relation $\bra{v_i} T^{2j} \ket{v_{i'}} = \delta_{i,j}e^{-kj}$ for $i,i'=1,2$, the matrix elements of $H-E_{\rm F}$ are obtained as
\begin{eqnarray}
\bra{v_1} H-E_{\rm F} \ket{v_1} & = & -\Delta^z(J_1+J_2) + J_2\Delta^{xy} \cos k, \\
\bra{v_2} H-E_{\rm F} \ket{v_1} & = & \frac{J_1\Delta^{xy}}{2}(1+\delta+(1-\delta)e^{-ik}), \\
\bra{v_1} H-E_{\rm F} \ket{v_2} & = & \bra{v_2} H-E_{\rm F} \ket{v_1}^{*}, \\
\bra{v_2} H-E_{\rm F} \ket{v_2} & = & \bra{v_1} H-E_{\rm F} \ket{v_1}.
\end{eqnarray}
This matrix has two eigenvalues 
\begin{eqnarray}
\varepsilon_{\pm,k}&=&-\Delta^z(J_1+J_2) + J_2\Delta^{xy} \cos k \nonumber \\
&& \pm \left| \frac{J_1\Delta^{xy}}{2}\right| \sqrt{ 2(1+\delta^2) + 2(1-\delta^2) \cos k}.
\end{eqnarray}
The spin gap $\Delta_{\mathrm{G}}$ can then be calculated as the lowest eigenvalue $\mathrm{min}_{k}\varepsilon_{-,k}$.
In addition to two trivial local minima,
\begin{align}
  \varepsilon_{-,k=0}=-\Delta^z(J_1+J_2) + J_2\Delta^{xy} - \left|J_1\Delta^{xy}\right|,
  \nonumber\\
  \varepsilon_{-,k=\pi}=-\Delta^z(J_1+J_2) - J_2\Delta^{xy} -\left|J_1\Delta^{xy}\delta\right|,
\end{align}
there may exist another local minimum at an incommensurate wave number $k_{\mathrm{inc}}$ in the thermodynamic limit. This solution can be obtained by differentiating $\varepsilon_{-,k}$ with respect to $k$ as 
\begin{eqnarray}
\partial_k E_{-} & = & -J_2\Delta^{xy} \sin k + \left| \frac{J_1\Delta^{xy} } {2} \right| (1-\delta^2) \sin k \nonumber \\
&& \times \left(  2(1+\delta^2) + 2(1-\delta^2) \cos k \right)^{-1/2},
\end{eqnarray}
leading to
\begin{eqnarray}
  |k_{\mathrm{inc}}| & = & \cos^{-1} \left[  \frac{J_1^2} {8J_2^2 } (1-\delta^2)  - \frac{  1+\delta^2 }{1-\delta^2} \right],
  \\
  \varepsilon_{-,\pm k_{\mathrm{inc}}} &=&-\Delta^z(J_1+J_2) \nonumber \\
  && - J_2\Delta^{xy}\frac{1+\delta^2}{1-\delta^2} -\frac{J_1^2\Delta^{xy}}{8J_2}(1-\delta^2),
\end{eqnarray}
in the case of $\delta^2\leq1$ and $-1 < \frac{J_1^2} {8J_2^2 } (1-\delta^2)  - \frac{  1+\delta^2 }{ 1-\delta^2 } < 1$.
The result $\Delta_{\mathrm{G}} = \min(\varepsilon_{-,0}, \varepsilon_{-,\pi}, \varepsilon_{-,\pm k_{\mathrm{inc}}})$ is plotted in the FPF phase of Fig.~\ref{fig:pd_xxz}(b).
In particular, for $\delta=0$, $\varepsilon_{-,-k_{\mathrm{inc}}}$ yeilds the spin gap $\Delta_{\mathrm{G}}$ for $-4<J_1/J_2<2$, while $\varepsilon_{-,\pi}$ for $J_1/J_2\leq-4$. 

\subsection{Spin gap in the UUDD phase in the Ising limit} \label{sec:u_phase}

Here, we assume that the number $L$ of sites in the system is a multiple of 4, $L=4n$, when the bond alternation $\delta$ is positive. 
The UUDD state in the Ising limit is written as
\begin{equation}
\ket{{\rm uudd}}=\otimes_{l=1}^{n} \ket{\uparrow \uparrow \downarrow \downarrow}. 
\end{equation}
The ground state energy under the periodic boundary condition reads
\begin{eqnarray}
E_{\rm uudd}& =& \bra{{\rm uudd}} H \ket{{\rm uudd}} \nonumber \\
& = & n \frac{J_1\Delta^z(2(1+\delta)-2(1-\delta))}{4} 
 - 4n \frac{J_2\Delta^z}{4} \nonumber \\
& = & n\Delta^z(J_1\delta-J_2).
\label{eq:E_UUDD}
\end{eqnarray}

In the case of $J_1/J_2 > -1$ and $\delta=0$, the lowest excited states within the $S^z_{\rm T}=1$ manifold are expressed as $\{\hat{T}^j \ket{\sigma^{\rm a}_{r}} \}^{0 \leq j < 4n}_{0 \leq r < n}$ with
\begin{equation}
\ket{\sigma^{\rm a}_{r}} = \ket{\uparrow \uparrow \downarrow} 
\otimes_{l=1}^{r} \ket{\uparrow \uparrow \downarrow \downarrow}
\otimes \ket{\uparrow} 
\otimes_{l'=1}^{n-r-1} \ket{\uparrow \uparrow \downarrow \downarrow}, 
\end{equation}
where $\hat{T}^j$ is the operator that translated the spins by $j$ sites.
Then, the spin gap $\Delta_{\rm G}$ is given by
\begin{equation}
\Delta_{\rm G} = \bra{\sigma^{\rm a}_{r}} H - E_{\rm uudd}\ket{\sigma^{\rm a}_{r}} = J_2 \Delta^z~.
\end{equation}

On the other hand, in the case of $-2 < J_1/J_2 < -1$ and $\delta=0$, the lowest excited states within the $S^z_{\rm T}=1$ manifold are given by $\{ \hat{T}^j \ket{\sigma^{\rm b}_{uvw}} \}$, where
\begin{eqnarray}
&& \ket{\sigma^{\rm b}_{uvw}} =  
\ket{\uparrow \uparrow \downarrow\downarrow \downarrow} 
\otimes_{l=1}^{u} \ket{\uparrow \uparrow \downarrow\downarrow} 
\otimes \ket{\uparrow} 
\otimes_{l'=1}^{v} \ket{\uparrow \uparrow \downarrow\downarrow} \nonumber \\
&& \otimes \ket{\uparrow} 
\otimes_{l''=1}^{w} \ket{\uparrow \uparrow \downarrow\downarrow} \otimes \ket{\uparrow} \otimes_{l'''=1}^{n-2-(u+v+w)} \ket{\uparrow \uparrow \downarrow\downarrow}
\end{eqnarray}
with $u\geq0  \land v\geq0 \land w\geq0 \land u+v+w \leq n-2$.
Then, the spin gap is given by
\begin{equation}
\Delta_{\rm G} = \bra{\sigma^{\rm b}_{uvw}} H - E_{\rm uudd}\ket{\sigma^{\rm b}_{uvw}} = 2J_2\Delta^z+J_1\Delta^z~.
\end{equation}

\begin{figure}[h]
\begin{center}
\includegraphics[width=8cm]{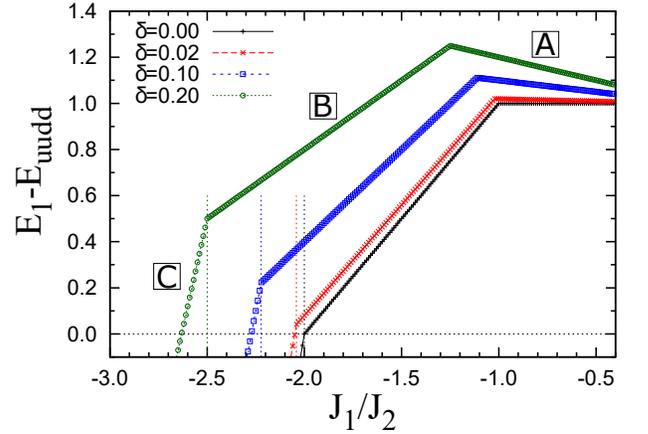}
\end{center} 
\caption{Energy difference between the lowest energy $E_1$ specified with $S^z_{\rm T} = 1$ and the energy $E_{{\rm uudd}}$ of up-up-down-down state for $\Delta^{xy}=0$ and $\Delta^z=1$ under the periodic boundary condition. Each vertical dotted line indicates the FPF--UUDD phase transition points for each $\delta$. The region A, B, and C are $J_1/J_2 > -1/(1-\delta)$, $-2/(1-\delta) < J_1/J_2 < -1/(1-\delta)$, and $J_1/J_2 < -2/(1-\delta)$, respectively. The UUDD state is a ground state in the region A and B, and the FPF state is a ground state in the region C. 
}
\label{fig:ene_ising}
\end{figure}

Turning on the positive $\delta$, the degeneracy of the two manifolds $\{ \hat{T}^{j}\ket{\sigma^{\rm a}_{r}}\}$ and $\{ \hat{T}^{j} \ket{\sigma^{\rm b}_{uvw}} \}$ is lifted. Then, the lowest-energy states within the $S_{\mathrm{T}}^z=1$ manifold are given by  $\{ \hat{T}^{2k} \ket{s^{\rm a}_{\ell}} \}^{0\leq k < 2n}_{\ell\in\{0,1\}}$ for $J_1/J_2 > -1/(1-\delta)$ [~region A in Fig.~\ref{fig:ene_ising}~]
with
\begin{equation}
\ket{s^{\rm a}_{0}} = \ket{\uparrow \uparrow \downarrow \uparrow} 
\otimes_{l=1}^{n-1} \ket{\uparrow \uparrow \downarrow \downarrow},
\end{equation}
\begin{equation}
\ket{s^{\rm a}_{1}}=
\ket{\uparrow \uparrow \uparrow \downarrow} 
\otimes_{l=1}^{n-1} \ket{\uparrow \uparrow \downarrow \downarrow}
\end{equation}
and $\{ \hat{T}^{2k} \ket{s^{\rm b}_{\ell p}} \}^{0\leq k < 2n}_{\ell\in\{0,1\}, 0 \leq p < n-1 }$ for $-2/(1-\delta) < J_1/J_2 < -1/(1-\delta)$ [~region B in Fig.~\ref{fig:ene_ising}~] with
\begin{equation}
\ket{s^{\rm b}_{0p}} =  
\ket{\uparrow \uparrow \downarrow \downarrow \downarrow \uparrow} 
\otimes_{l=1}^{p} \ket{\uparrow \uparrow \downarrow \downarrow}
\otimes \ket{\uparrow \uparrow} 
\otimes_{l'=1}^{n-2-p} \ket{\uparrow \uparrow \downarrow \downarrow}
,
\end{equation}
\begin{equation}
\ket{s^{\rm b}_{1p}} = \ket{\uparrow \uparrow \uparrow \downarrow \downarrow \downarrow}
\otimes_{l=1}^{p} \ket{\uparrow \uparrow \downarrow \downarrow}
\otimes \ket{\uparrow \uparrow} 
\otimes_{l'=1}^{n-2-p} \ket{\uparrow \uparrow \downarrow \downarrow},
\end{equation}
respectively.
Then, the spin gap is given by
\begin{equation}
\Delta_{\rm G} = \left \{ 
\begin{matrix}
J_2 \Delta^z - J_1 \delta \Delta^z & [ \frac{J_1}{J_2} > -\frac{1}{1-\delta} ] \\
2J_2\Delta^z + J_1(1-2\delta)\Delta^z & [ -\frac{2}{1-\delta} < \frac{J_1}{J_2} < -\frac{1}{1-\delta} ]
\end{matrix}
\right.~.
\end{equation}
Thus, the effect of finite $\delta$ 
introduces a non-monotonic behavior of the spin gap with respect to $J_1/J_2$, and the result of $\Delta_{\mathrm{G}}$ is plotted in Fig.~\ref{fig:pd_xxz}(b).

\subsection{Effective Hamiltonian in the strongly dimerized case}\label{sec:effective_s1}

Here, to understand the nature of the phase transitions among the UUDD phase, D$_\pm$ phase and VCD$_\pm$ phases, we consider a mapping to an effective Hamiltonian in the strongly dimerized limit of $(1-|\delta|) \ll J_2/|J_1| \ll (1+|\delta|)$. The following analysis holds in the strongly dimerized case, i.e., $|\delta| \to 1$, but the nature of the mapping and the associated phases and transitions should also hold for weakly dimerized case as long as the phases are adiabatically connected from strong to weakly dimerized regimes. 

In the strongly dimerized limit, it is reasonable to perform the following projection onto the effective (vector-chiral) spin-1 degree of freedom formed on the strong nearest-neighbor ferromagnetic bond:
\begin{equation}
p_\theta = \ket{\uparrow\uparrow}\bra{\uparrow\uparrow} + \ket{0,\theta}\bra{0,\theta} + \ket{\downarrow\downarrow}\bra{\downarrow\downarrow}
\end{equation}
with $\ket{0,\theta} \equiv (e^{-i\theta/2} \ket{\uparrow\downarrow}+e^{i\theta/2}\ket{\downarrow\uparrow})/\sqrt{2}$, which possess the local vector chirality,
\begin{equation}
\bra{0,\theta} (S^x \otimes S^y - S^y \otimes S^x) \ket{0,\theta}=-\frac{1}{2}\sin \theta,
\end{equation}
where $\theta$ is a parameter describing the ground state, and is 0 in the $D_\pm$ and UUDD phases while nonzero ($0<|\theta|\le\pi/2$) in the VCD$_\pm$ phases.
Then, pairs of spin-1/2 operators are projected onto effective spin-1 operators as
\begin{eqnarray}
p_\theta (S^x \otimes S^x + S^y \otimes S^y) p_\theta^{\dagger} & = & \frac{1}{2}\left(1 - \left( s^z \right)^2\right)\cos \theta, \\ \label{eq:p_th_1}
p_\theta (S^z \otimes S^z) p_\theta^{\dagger} & = & \frac{1}{2}\left( s^z \right)^2 - \frac{1}{4}, 
\end{eqnarray}
\begin{eqnarray}
p^{~}_\theta ( \1 \otimes S^x) p_\theta^{\dagger} & = & \frac{1}{2}\left(s^x\cos\frac{\theta}{2} + s^y\sin \frac{\theta}{2}\right),  \\
p^{~}_\theta ( \1 \otimes S^y) p_\theta^{\dagger} & = & \frac{1}{2}\left(-s^x\sin\frac{\theta}{2}  + s^y\cos \frac{\theta}{2}\right),  \\
p^{~}_\theta ( \1 \otimes S^z) p_\theta^{\dagger} & = & \frac{1}{2}s^{z}, 
\end{eqnarray}
\begin{eqnarray}
p^{~}_\theta ( S^x \otimes \1) p_\theta^{\dagger} & = & \frac{1}{2}\left(s^x\cos\frac{\theta}{2} - s^y\sin \frac{\theta}{2}\right),  \\
p^{~}_\theta ( S^y \otimes \1) p_\theta^{\dagger} & = & \frac{1}{2}\left(s^x\sin\frac{\theta}{2} + s^y\cos \frac{\theta}{2}\right),  \\
p_\theta ( S^z \otimes \1) p_\theta^{\dagger} & = & \frac{1}{2}s^{z}. \label{eq:p_th_8}
\end{eqnarray}
Using the relations given in Eqs.~(\ref{eq:p_th_1})-(\ref{eq:p_th_8}), the effective spin-1 Hamiltonian $H_\theta =P_{\theta}HP^{\dagger}_\theta$ with $P_\theta \equiv \otimes_i p^{~}_\theta$ is obtained as
\begin{eqnarray}
H_\theta & = & \sum_j J^{xy}_{\theta} (s^x_j s^x_{j+1} + s^y_j s^y_{j+1}) + J^{z}_{\theta} s^z_j s^z_{j+1} \nonumber \\
&& + d_\theta (s^x_j s^y_{j+1}-s^y_j s^x_{j+1}) + D_{\theta}(s^z_j)^2 + C_{\theta}
\label{eq:h_phi}
\end{eqnarray}
with 
\begin{eqnarray}
J^{xy}_\theta & = & \frac{\Delta^{xy}}{4}(J_1(1-|\delta|)\cos\theta+2J_2), \\
J^{z}_\theta & = & \frac{\Delta^{z}}{4}(J_1(1-|\delta|)+2J_2), \\
d_\theta & = & -\frac{\Delta^{xy}}{4} J_1(1+|\delta|) \sin \theta, \\
D_\theta & = & J_1(1+|\delta|)\frac{\Delta^z-\Delta^{xy}\cos \theta}{2}, \\
C_\theta & = & J_1(1+|\delta|)\frac{2\Delta^{xy}\cos \theta-\Delta^z}{4}. 
\end{eqnarray}
The Dzyaloshinskii-Moriya interaction term with the coupling constant $d_\theta$ can be gauged away from the Hamiltonian (\ref{eq:h_phi}) by considering the following spin axes rotations, 
\begin{eqnarray}
s^x_j \rightarrow \tilde{s}^x_j & = & s^x_j \cos \theta_j + s^y_j \sin \theta_j, \nonumber \\
s^y_j \rightarrow \tilde{s}^y_j & = & -s^x_j \sin \theta_j + s^y_j \cos \theta_j, \nonumber \\
s^z_j \rightarrow \tilde{s}^z_j & = & s^z_j, \nonumber \\
\end{eqnarray}
with
\begin{equation}
\theta_j=(j-1) \varphi,~\tan \varphi = d_\theta/J^{xy}_\theta.
\end{equation}
Then, the Hamiltonian Eq.~(\ref{eq:h_phi}) is reduced to the effective spin-1 Hamiltonian $\tilde{H}_\theta$, which is equivalent to the $S=1$ XXZ chain with the single-ion anisotropy,
\begin{eqnarray}
\tilde{H}_\theta & = & \sum_j \tilde{J}^{xy}_{\theta} (\tilde{s}^x_j \tilde{s}^x_{j+1} + \tilde{s}^y_j \tilde{s}^y_{j+1}) + J^{z}_{\theta} \tilde{s}^z_j \tilde{s}^z_{j+1} \nonumber \\
&& + D_{\theta}(\tilde{s}^z_j)^2 + C_{\theta} 
\label{eq:tilde_h_phi}
\end{eqnarray}
with
\begin{equation}
\tilde{J}^{xy}_{\theta} = {\rm sgn}(J^{xy}_\theta) \sqrt{(J^{xy}_\theta)^2+d^2_\theta}. 
\end{equation}

The global phase diagram of $\tilde{H}_\theta$ for $\tilde{J}^{xy}>0$ has already been investigated by the numerical exact diagonalization with finite-size scaling analyses~\cite{chen_prb_2003}. It contains the N\'{e}el, Haldane, and Large-$D$ phases for $\tilde{J}^{xy}_{\theta} > 0 \land J^{z}_{\theta} > 0$. These phases corresponds to the UUDD, D$_+$ (VCD$_+$) and D$_-$ (VCD$_-$) phases for $\theta=0$ ($\theta>0$), respectively, in the original spin-1/2 Hamiltonian $H$. Then, it is reported that the Haldane--N\'{e}el phase transition and the Haldane--Large-$D$ phase transition are continuous phase transitions that belong to the Ising universality class (the central charge $c=1/2$) and to the Gaussian universality class ($c=1$), respectively~\cite{chen_prb_2003}. Indeed, our current numerical finding on the Ising universality class for the D$_+$--UUDD phase transition on the easy-axis side of the original spin-$1/2$ Hamiltonian$H$ [Sec.~\ref{sec:transitions:D_+-UUDD}] agrees with that of the Haldane--N\'{e}el phase transition for $\tilde{H}_\theta$ with $\theta=0$. Furthermore, our previous numerical finding on the Gaussian universality class for the VCD$_+$--VCD$_-$ phase transition in the easy-plane side of $H$~\cite{UO14} can also be elucidated from that of the Haldane--large-$D$ phase transition for $\tilde{H}_\theta$ with finite $\theta$.
The D$_+$--D$_-$ phase transition that most likely appears for small $|J_1/J_2|$ in the case of large $|\delta|$ in the spin-1/2 Hamiltonian $H$ with the easy-plane anisotropy should also belong to the Gaussian universality class from the analogy to the Haldane--Large-$D$ phase transition for $\tilde{H}_\theta$ with $\theta=0$.

\section{Discussion and conclusions}
\label{sec:conclusions}

  We have investigated the ground-state properties and the spin gap of the spin-$1/2$ frustrated spin XXZ chain with the ferromagnetic first-neighbor coupling $J_1$ and the antiferromagnetic second-neighbor coupling $J_2$, both with and without an alternation in $J_1$. Using the iTEBD and iDMRG methods, we have numerically completed the ground-state phase diagram as well as the maps of the spin gap, the wave number of the maximum spin correlation, and the transverse magnetic susceptibility on both the easy-plane and easy-axis sides.   We have also derived analytic expressions for the spin gap in the fully polarized ferromagnetic phase (FPF) and in the up-up-down-down (UUDD) phase near and in the Ising-limit, respectively, in the case with a finite bond alternation $\delta$ in the first-neighbor exchange coupling. These maps, summarized in Fig.~\ref{fig:pd_xxz} and Table~\ref{table:pd_xxz}, will be useful for understanding low-temperature experimental results on the basis of a single frustrated $J_1$-$J_2$ spin-$1/2$ chain.

  In particular, our numerical results has for the first time uncovered the anisotropic dimer (D$_-$) phase on the side of easy-axis magnetic anisotropy. We have shown that this phase is surrounded by the partially polarized ferromagnetic (PPF) phase, the up-up-down-down (UUDD) phase, and the (nearly) isotropic dimer phase (D$_+$). Clear first-order phase transitions have been found at the D$_-$--PPF and PPF--UUDD phase. The D$_-$--UUDD phase transition either is weakly first-order or belongs to the Ising universality class. The D$_-$--D$_+$ phase transitions is of the second order and belongs to the Gaussian universality class with the central charge $c=1$. The nature of D$_-$--UUDD and D$_-$--D$_+$ phase transitions has also been explained as the Haldane--Large-$D$ and Haldane--N\'{e}el phase transitions, respectively, by mapping the original spin-$1/2$ model in the strongly dimerized case onto an effective spin-1 XXZ chain model with single-ion anisotropy $D$. Note that the D$_-$--D$_+$ phase transition is a symmetry-protected topological phase transition protected by the time-reversal, bond-inversion, and $Z_2 \times Z_2$ symmetries. This analysis for obtaining the Gaussian universality class holds for the VCD$_-$--VCD$_+$ phase transition on the side of easy-plane exchange magnetic anisotropy, which solves the issue posed from previous numerical finding~\cite{UO14}.

Our results have crucial implications to experiments on low-temperature magnetic properties of quasi-one-dimensional spin-$1/2$ compounds.
Large exchange magnetic anisotropy on both the easy-plane and easy-axis sides can yield an energy gap in the spin excitation above a nonmagnetic ground state in the D$_-$ and VCD$_-$ phases. However, the ground state necessarily possesses a moderately large transverse magnetic susceptibility. This feature is robust and also holds even when the U(1) spin symmetry is absent~\cite{note_XYZ}. It may happen in experiments that while the magnetic susceptibility is strongly suppressed as in the D$_+$ and VCD$_+$ phases, the spin gap is too large for the D$_+$ and VCD$_+$ phases. 
Such case can hardly be explained within single frustrated spin-$1/2$ models, but will demand interchain interaction that enhances the spin gap, as we will show below that this is the case for Rb$_2$Cu$_2$Mo$_3$O$_{12}$.

Now, we explain the extent to which single $J_1$-$J_2$ spin-$1/2$ chain models can and cannot explain experimental findings of the quasi-1D spin-$1/2$ chain compound Rb$_2$Cu$_2$Mo$_3$O$_{12}$ that hosts a nonmagnetic ground state~\cite{UO18}.
It is reasonable to start from the SU(2) symmetric model with no bond alternation $\delta=0$ and then to consider effects of perturbations.
The wave number $q_{\mathrm{max}} \sim \pi/4$, which corresponds to the eight-spin periodicity, of the maximum spin correlation has been observed with inelastic neutron-scattering experiments. From this value, the ratio $J_1/J_2$ is estimated as $-3.6$. Then, the magnetic susceptibility of the powder samples can be fit in the temperature range from 300~K down to 40~K by taking $|J_1|=0.62$~meV and $J_2=0.17$~meV. The first spin excitation energy has also been measured as $\sim0.2$~meV in inelastic neutron-scattering experiments. This energy gap is too large for the spin gap in the D$_+$ phase for $\delta=0$, since $\Delta_{\mathrm{G}}<0.03J_2$ in the D$_+$ phase with and without easy-plane anisotropy. Attributing this large energy gap to the bond alternation demands a rather large value $\delta=0.1$. However, with this value of $\delta$, $q_{\mathrm{max}}$ approximates to $\pi/2$, in stark contrast to the neutron result of $q_{\mathrm{max}} \sim \pi/4$. There is another way to enhance the energy gap while keeping the relation $q_{\mathrm{max}}\sim \pi/4$ intact. Increasing easy-plane anisotropy, namely, decreasing $\Delta^z$, VCD$_\pm$ and D$_-$ phases appear. The VCD$_+$ phase is also excluded since the spin gap is too small as in the D$_+$ phase. In the VCD$_-$ and D$_-$ phases, the spin gap may increase up to $\Delta_{\rm G}= 0.93J_2$, for instance, for $(J_1/J_2,\Delta^{z},\delta)=(-2.02,0.0,0.2)$. Thus, if we take $J_2=0.21$ meV, we reproduce $\Delta_{\rm G}=0.2$ meV keeping the wave number $q_{\mathrm{max}}\sim\pi/4$. However, the transverse magnetic susceptibility is as large as 0.7 $\mu_{\rm B}$/Cu$\cdot$T in the ground state. This contradicts to experimental observations of a suppression of the magnetic susceptibility to 0.005 $\mu_{\rm B}$/Cu$\cdot$T below 0.08~K and the associated magnetization curve that starts to rise steeply only at $2$~T. 
In general, Dzyaloshinskii-Moriya interactions as well as XYZ exchange anisotropy $\Delta^x=\Delta^{xy}<\Delta^y<\Delta^z$ do not increase the first spin excitation energy as long as the spin gap is already finite in the case of $\Delta^x=\Delta^y=\Delta^{xy}$, and thus should not be responsible for enhancing the size of the spin gap. 
Thus, we are naturally led to the next candidate, namely, an unfrustrated interchain interaction within a pair of $J_1$-$J_2$ spin-$1/2$ chains. It has been shown that it can enhance the spin gap while keeping $q_{\mathrm{max}}$ intact. 
Indeed, the two-leg ladder model of a pair of $J_1$-$J_2$ frustrated spin-$1/2$ chains that are coupled by an antiferromagnetic rung interaction $J'$ can quantitatively explain overall experimental findings. Namely, choosing $J_1=-9.82$~meV, $J_2=3.03$~meV, and $J'=1.77$~meV, and taking rather large Dzaloshinski-Moriya interactions of the order 3.8~meV, the model accounts all of  the magnetic susceptibility, the magnetic field strength for closing the spin gap, and inelastic neutron-scattering and electron  spin resonance spectra~\cite{UO18}. 

Lastly, we briefly mention the case of Cs$_2$Cu$_2$Mo$_3$O$_{12}$~\cite{Hase05,Fujimura16,Goto17,Yagi18}. 
Recent neutron diffraction experiments on this material have shown that the spins order ferromagnetically within single chains and antiferromagnetically between the two coupled chains~\cite{Yasui20}. It is naturally expected that the effective single spin-$1/2$ chain model lies in the ferromagnetic phases on the easy-axis side or the TLL phase, which shows a quasi-long-range ferromagnetic order, on the easy-plane side. As in Rb$_2$Cu$_2$Mo$_3$O$_{12}$, the two adjacent chains should be coupled antiferromagnetically.  Actually, this rung interaction does not allow the full spin polarization, since the fully spin polarized N\'eel state cannot be an eigenstate of the ladder Hamiltonian. If we take a naive estimate of the ratio $J_1/J_2$ to be $-2.8$~\cite{Hase05}, it is likely that the compound possesses easy-axis exchange anisotropy, since with easy-plane anisotropy, the single-chain model shows the vector-chiral order, which is readily driven to a long-range spiral magnetic order by three-dimensional interactions of the order of the N\'eel temperature 1.85~K~\cite{Fujimura16}. 
It is also possible that the compound has a moderately large $J'$ as in Rb$_2$Cu$_2$Mo$_3$O$_{12}$ and a larger  $|J_1|/J_2$ than 2.8.
Inelastic neutron-scattering experiments are required for further quantitative theoretical analyses.

\acknowledgments
The authors acknowledge S. Furukawa for discussions and Y. Yasui, M. Hagiwara, T. Masuda for collaborations and stimulating discussions on Rb$_2$Cu$_2$Mo$_3$O$_{12}$. 
HU thanks to S. Yunoki for continuous encouragement and is  supported by KAKENHI No. 17K14359, 17H02926, and by JST PRESTO No. JPMJPR1911. SO is supported by KAKENHI No. 16K05426 and 19K03729 by the RIKEN iTHES project. A part of numerical calculations was performed by using the HOKUSAI-Great Wave supercomputing system at RIKEN and by the K computer provided by the RIKEN Center for Computational Science.

\appendix
\section{Two parameter scaling for numerical error of iTEBD calculation}
\label{sec:itebd}
\begin{figure}[h]
\begin{center}
\includegraphics[width=8cm]{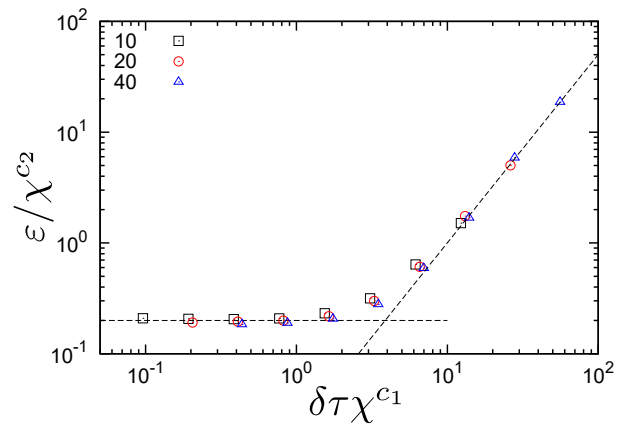}
\end{center}
\caption{Two parameter scaling for $\varepsilon$ and $\delta \tau$ with respect to $\chi$ up to 40 for $\mathcal{H}_{\mathrm{XXZ}}$ with $J_1/J_2=-5/3$ and $\Delta^{xy}=\Delta^z=1$. Scaling parameters $c_1=1.09\pm0.03$ and $c_2=-1.87 \pm 0.06$ are estimated by the Bayesian inference method~\cite{Harada2011}.}
\label{fig:err_vs_dtau_j1j2a06}
\end{figure}

There are two origins of calculation errors in the iTEBD method~\cite{Vidal07}. One is the step size $\delta \tau$ of the imaginary time in the Suzuki-Trotter (ST) decomposition~\cite{Trotter1958,Suzuki1976}. The other is the finite bond dimension $\chi$ of the matrix-product state (MPS). In the limit where $\chi \rightarrow \infty$ with $\delta\tau$ being fixed, the dominant factor of the error is the time step size $\delta \tau$ of the ST decomposition. 
In this limit, the error of the total energy
$\varepsilon(\delta \tau,\chi)=E(\delta \tau,\chi)-E_{\rm exact}$, where $E(\delta \tau,\chi)$ is the variational
energy of iTEBD calculation and $E_{\rm exact}$ is the
exact ground state energy,
behaves as
\begin{equation}
\varepsilon(\delta \tau, \chi \rightarrow \infty) \sim (\delta \tau)^{n+1}
\label{eq:e1}
\end{equation} 
with $n$ being the order of the ST decomposition.

On the other hand, in the limit of $\delta \tau \rightarrow 0$ with $\chi$ being fixed, the leading factor of the error becomes finite-$\chi$ effects in MPS. From the finite-entanglement scaling~\cite{Pirvu2012}, we expect that the asymptotic form of the error becomes 
\begin{equation}
\varepsilon(\delta \tau \rightarrow 0, \chi) \sim \chi^{c_2}, 
\label{eq:e2}
\end{equation}
where we assume the real correlation length is extremely large, compared to a controllable length scale introduced by the finite-$\chi$ effects.  

Here, we propose a scaling hypothesis 
\begin{equation}
\varepsilon(\delta \tau, \chi) = \chi^{c_2} F(\delta \tau \chi^{c_1})
\label{eq:scaling_hypothesis}
\end{equation}
to satisfy both asymptotic form Eq.~(\ref{eq:e1}) and Eq.~(\ref{eq:e2}), where the function $F$ has the asymptotes,
\begin{equation}
F(\delta \tau \chi^{c_1}) \sim \left\{
\begin{matrix}
\chi^{-c_2} (\delta \tau)^{n+1} & ,&  \delta \tau \chi^{c_2} \gg 1 \\
\mathrm{const.} & , & \delta \tau \chi^{c_1} \ll 1 \\
\end{matrix}
\right. ,
\end{equation}
and apply this scaling analysis to the energy error for $\mathcal{H}_{\mathrm{XXZ}}$ in Eq.~(\ref{eq:H_0_XXZ}) with $J_1/J_2=-5/3$ and $\Delta^{xy}=\Delta^z=1$~\cite{Agrapidis2017, Efthimia2019}.
Then, we confirm that the energy error $\varepsilon$ with several $\chi$ are nicely on the universal function with appropriate $c_1$ and $c_2$ and employ a condition $\delta \tau \chi^{c_1} \sim 4$, where two asymptotic lines intersect. This provides a condition for efficient calculation of iTEBD, and the suitable step size is found to be $\delta \tau \sim 0.008/J_2$ for $\chi=300$. Under the conditions, the error of the ground-state energy per site in the Heisenberg case ($\Delta^{xy}=\Delta^z=1$) for $J_1/J_2=-5/3$ is found to be $\sim5\times 10^{-6} J_2$. The condition of $\delta \tau \chi^{c_1}$, where two asymptotic lines intersect, depends on the parameter set. However, it is difficult to perform the scaling analysis in the all of parameter space because we should refer to the numerically exact ground state energy, that is quite time consuming process. Therefore, the constant step size $\delta \tau = 0.008/J_2$ is employed in the all of iTEBD calculation, and we check the convergence of the calculation with up to $\chi=300$.  

\section{Convergence of order parameters with respect to bond dimensions in iTEBD}
\label{sec:itebd_m}
\begin{figure}[h]
\begin{center}
\includegraphics[width=8cm]{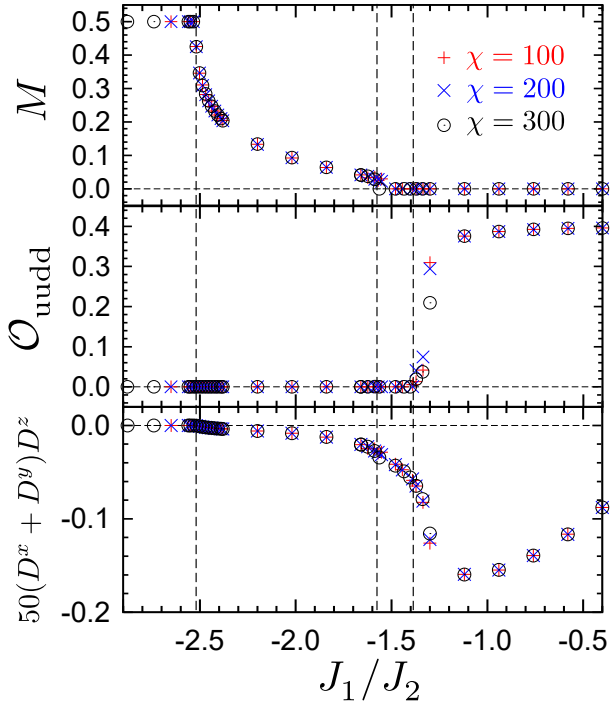}
\end{center}
\caption{Example of $\chi$ dependence of the order parameters for $\Delta^{xy}_{~}=0.45$, $\Delta^{z}_{~}=1$, and $\delta=0.02$ with $\chi=100$, $200$, and $300$. The vertical broken lines are phase boundary shown in Fig.~\ref{fig:pd_xxz}(a) or Fig.~\ref{fig:PF_D-_U}(c).}
\label{fig:O_m}
\end{figure}

In this article, we estimate all of order parameters by use of the iTEBD method~\cite{Vidal07} up to $\chi=300$. The order parameters, of course, depend the value of $\chi$. In Fig.~\ref{fig:O_m}, we show an example of the convergence behaviors, with respect to $\chi$, of $M$ in Eq.~(\ref{eq:M}), $\mathcal{O}^{~}_{\rm uudd}$ in Eq.~(\ref{eq:uudd}), and $(D^x_{~}+D^y_{~})D^z_{~}$ with Eq.~(\ref{eq:D}) for the parameter set $\Delta^{xy}_{~}=0.45$, $\Delta^{z}_{~}=1$, and $\delta=0.02$, for which three successive phase transitions occur from the FPF phase through the PPF and D$_-$ phases to the UUDD phase. The order parameters converge with respect to $\chi$ except near phase boundaries, and the error of phase-transition points with respect to $J_1/J_2$ are smaller than 0.1 in the example. Therefore, we plot the phase boundaries estimated by the calculation with $\chi=300$ in Fig.~\ref{fig:pd_xxz}(a). 

\section{Dependence of spin gap on the system size and the bond dimensions in iDMRG}
\label{sec:idmrg}
\begin{figure}[h]
\begin{center}
\includegraphics[width=8cm]{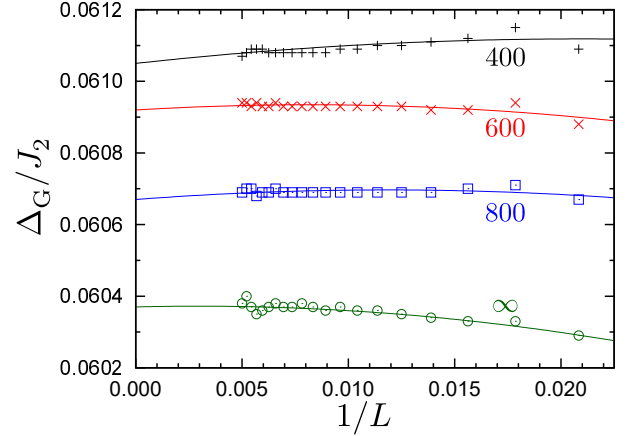}
\end{center}
\caption{Example of a finite size scaling of the spin gap for $J^{~}_1/J^{~}_2=-2.56$, $	\Delta^{xy}_{~}=0.9$, $\Delta^{z}_{~}=1$, and $\delta=0.2$ with $\chi=400$, $600$, and $800$. The data of $\chi=\infty$ is estimated by the linear fitting with respect to $1/\chi$.   
}
\label{fig:gap_m}
\end{figure}

Here, we explain the finite-size scaling analysis of the spin gap $\Delta^{~}_{\rm G}/J^{~}_{2}$ in the global phase diagram [Fig.~\ref{fig:pd_xxz} (b)], except for the FPF and PPF phases. We performed the iDMRG~\cite{white92,white93,iDMRG} calculations, exploiting the U(1) symmetry, and obtained the lowest energies for $S^z_{\rm tot}=0$ and $1$ as functions of $L$ and estimate the energy gap between them. Then, we perform the second-order polynomial fittings as functions of $1/L$ with $L$ up to 200. We checked the dependence of the gap on $\chi$ and estimate the difference between $\Delta^{~}_{\rm G}/J^{~}_{2}$ with $\chi=800$ and $\chi=\infty$. 
Figure \ref{fig:gap_m} demonstrates the analysis in the particular case of $J^{~}_1/J^{~}_2=-2.56$, $	\Delta^{xy}_{~}=0.9$, $\Delta^{z}_{~}=1$, and $\delta=0.2$. 
The difference in Fig.~\ref{fig:gap_m} is about 0.0003 and is negligibly small on the scale of Fig.~\ref{fig:pd_xxz}(b). Also, we have confirmed that the $\chi$ dependence safely converges within 10\% error when $\Delta^{~}_{\rm G}/J^{~}_{2} > 0.03$ in our calculations. Thus, we plot $\Delta^{~}_{\rm G}/J^{~}_{2}$ with $\chi=800$ in the figure.

\clearpage

\bibliographystyle{apsrev4-1}

\begin{thebibliography}{80}
\bibitem [{not()}]{note_XYZ}%
  \BibitemOpen
  \href@noop {} {}\bibinfo {note} {To suppress the large $x$ and $y$ components
  of magnetic susceptibility simultaneously, a recovery of the SU(2) symmetry
  is required in the ground state. The anisotropy $\Delta^x \neq \Delta^y$
  cannot contribute to this symmetry recovery.}\BibitemShut {Stop}%
\end{thebibliography}
%

\end{document}